\documentclass{aa}
\usepackage{graphicx}
\begin{document}
   \title{Constraints on stellar convection from multi-colour photometry of $\delta$ Scuti stars}

   \author{J. Daszy\'nska-Daszkiewicz$^{1,2}$, W. A. Dziembowski$^{2,3}$, A. A.
   Pamyatnykh$^{2,4}$}

   \offprints{J. Daszy\'nska-Daszkiewicz, \email{daszynska@astro.uni.wroc.pl}}

   \institute{{1} Astronomical Institute of the Wroc{\l}aw University,
   ul. Kopernika 11, 51-622 Wroc{\l}aw, Poland\\
   {2} Copernicus Astronomical Center, Bartycka 18, 00-716 Warsaw, Poland\\
   {3} Warsaw University Observatory, Al. Ujazdowskie 4, 00-478 Warsaw,
Poland\\
   {4} Institute of Astronomy, Russian Academy of Sciences,
   Pyatnitskaya Str. 48, 109017 Moscow, Russia\\
    }
   \date{Received ...; accepted ...}
   \abstract{
   In $\delta$ Scuti star models, the calculated amplitude ratios
   and phase differences for multi-colour photometry exhibit a strong
   dependence on convection. These observables are tools for determination of
   the spherical harmonic degree, $\ell$, of the excited modes.
   The dependence on convection enters through the complex parameter $f$,
   which describes bolometric flux perturbation.
   We present a method of simultaneous determination of $f$
   and harmonic degree $\ell$  from  multi-colour data and
   apply it to three $\delta$ Scuti stars. The method indeed works.
   Determination of $\ell$ appears unique and the inferred $f$'s
   are sufficiently accurate to yield a useful constraint on models
   of stellar convection. Furthermore, the method helps to refine stellar
   parameters, especially if the identified mode is radial.
\\
   \keywords{stars: $\delta$ Scuti variables --
             stars: oscillation --
             stars: convection
          }
}

   \titlerunning{Constrains on stellar convection from multi-colour photometry}
   \authorrunning{Daszy\'nska-Daszkiewicz et al.}
   \maketitle

\section{Introduction}

The $\delta$~Scuti stars are pulsating variables located
in the HR diagram at the intersection of the classical
instability strip with the main sequence and somewhat above it.
The observables of primary interest for asteroseismology are
oscillation frequencies. However, information about amplitudes and
phases of oscillations in various photometric passbands is also useful.
So far the main application of multi-colour photometry of $\delta$ Scuti
stars has been determination of $\ell$ degree of observed
modes (see e.g. Balona \& Evers 1999, Garrido 2000).
This is an important application
because knowledge of $\ell$ is an essential step for mode
identification. The $\ell$ diagnostic makes use of diagrams in which the
amplitude ratio determined in two passbands is plotted against the
corresponding phase difference. The observational data are
compared with ranges calculated for relevant stellar models and
assumed $\ell$ values. The ranges reflect uncertainties in stellar
parameters and physics. The inference is easy if the ranges do not
overlap. This is largely true for $\beta$ Cephei stars, but not
for $\delta$ Scuti stars. For the latter objects, a major
uncertainty in the calculated ranges arises from lack of adequate
theory of stellar convection.

Calculations of the amplitude ratios and phase differences make
use of the complex parameter $f$, which gives the ratio of the
radiative flux perturbation to the radial displacement at
photosphere. The parameter is obtained with the linear
nonadiabatic calculations of stellar oscillations. The
problem, which has been already emphasized by Balona \& Evers
(1999), is that $f$ is very sensitive to convection whose
treatment still remains rather uncertain.

The strong sensitivity of calculated mode positions in the
diagnostic diagrams to treatment of convection is not necessarily
a bad news. Having data from more than two passbands we may try to determine
simultaneously $\ell$ and $f$. If we succeed, the $f$-value
inferred from the data would yield then a valuable constraint on
models of stellar convection. The aim of our work is to examine
prospect for extracting $f$ from multi-colour photometry of
$\delta$ Scuti stars.

  \begin{figure*}
  \centering
    \includegraphics[bb=24 375 537 733, width=17cm,clip]{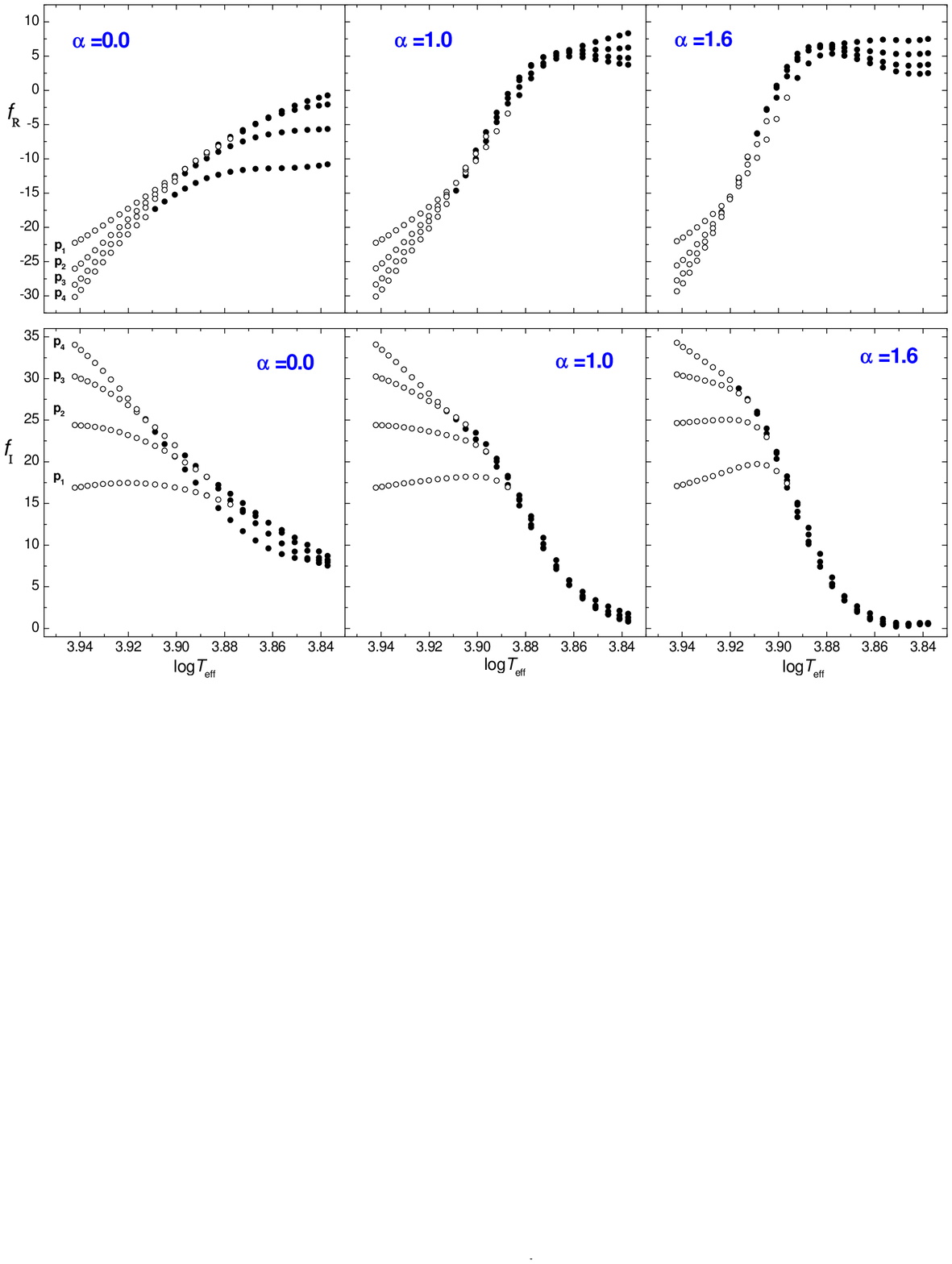}
           \caption{ The real and imaginary part of the $f$-parameter for
             radial oscillation of a 1.9 $M_\odot$ star in the main-sequence phase.
             In the panels from left to right we show the $f$-parameter for three
             values of $\alpha$.   Dots represent the unstable
             modes, whereas open circles the stable ones.}
\label{aaaaaa}
\end{figure*}

In the next section we demonstrate strong sensitivity of $f$ and,
in a consequence, of mode position in the diagnostic diagrams, to
the description of the convective flux. Our treatment of
convection is very simplistic. We rely on the mixing-length theory
and the convective flux freezing approximation. We study how
calculated positions in the diagnostic diagrams vary with the
changes of mixing-length parameter, $\alpha$.

Our method of inferring $f$ from the data is described in Sect.\,3.
The subsequent section presents application of the method to
several $\delta$ Scuti stars. In the last section we summarize
results of our analysis.

\section{Calculated mode positions in the diagnostic diagrams}

We use here the standard description of oscillating stellar
photospheres (cf. Cugier et al. 1994). The local displacement is
adopted in the form
$$\delta r(R,\theta,\varphi)=\varepsilon R {\rm Re}\{ Y_\ell^m
{\rm e}^{-{\rm i}\omega t}\}, $$
where $\varepsilon$ is a small complex parameter fixing mode
amplitude and phase. The associated perturbation of the bolometric flux,
${\cal F}_{\rm bol}$, and the local gravity, $g$, are then given by
$$\frac{ \delta {\cal F}_{\rm bol} } { {\cal F}_{\rm bol} }= \varepsilon
{\rm Re}\{ f Y_\ell^m {\rm e} ^{-{\rm i} \omega t} \},$$
and
$$\frac{\delta g}{g} = - \left( 2 +
\frac{\omega^2 R^3}{G M} \right) \frac{\delta R}{R}.$$
With the static plane-parallel approximation for the atmosphere we
can express the complex amplitude of the monochromatic flux
variation as follows (see e.g. Daszy\'nska-Daszkiewicz et al.
2002)
$$A^{\lambda}(i) = \varepsilon Y_\ell^m(i,0) b_\ell^\lambda (
D_{1,\ell}^{\lambda} +D_{2,\ell} +D_{3,\ell}^{\lambda}),
\eqno(1)$$
where $i$ is the inclination angle and $\lambda$ identifies the passband,
$$D_{1,\ell}^{\lambda} = \frac14  f \frac{\partial \log ( {\cal
F}_\lambda |b_{\ell}^{\lambda}| ) } {\partial\log T_{\rm{eff}}},
$$
$$D_{2,\ell} = (2+\ell )(1-\ell ),$$
$$D_{3,\ell}^{\lambda}= -\left( \frac{\omega^2 R^3}{G M}
 + 2 \right) \frac{\partial \log ( {\cal F}_\lambda
|b_{\ell}^{\lambda}| ) }{\partial\log g}.$$
The disc averaging factor, $b_{\ell}$, is defined by the integral
$$b_{\ell}^{\lambda}=\int_0^1 h_\lambda(\mu) \mu P_{\ell}(\mu) d\mu,$$
where the function $h_{\lambda}$ describes the limb darkening
law and $P_{\ell}$ is the Legendre polynomial.
The partial derivatives of ${\cal F}_\lambda |b_{\ell}^{\lambda}|$
may be  calculated numerically from tabular data. Here we rely on
Kurucz (1998) models and Claret (2000) computations of limb
darkening coefficients. With Eq. 1 we can directly obtain the
amplitude ratio and the phase difference for chosen pair of
passbands, which are called nonadiabatic observables.
Nonadiabaticity of oscillations enters through the complex
parameter $f$, which is the central quantity of our paper.

In Fig.\,1 we illustrate how the choice of the mixing-length
parameter, $\alpha$, affects the value of $f$. We can see the
large effect of the choice, particularly between $\alpha=0.0$ and
$\alpha=1.0$ in the cooler part of the sequence where all modes
are unstable. We have found that important is only the value of
$\alpha$ in the H ionization zone. Models calculated with $\alpha$
fixed in this zone and varied in the HeII ionization zone yield
very similar values of $f$.

In the next two figures we show how the differences in $\alpha$
are reflected in the $A_{b-y}/A_y~vs.~\varphi_{b-y}-\varphi_y$
diagram employing the Str\"omgren passbands . The effect of
varying $\alpha$ is very large indeed. First, in Fig.\,2 we show
all the unstable modes in the frequency range covering $p_1$ to
$p_4$ radial mode. The domains of different $\ell$'s partially
overlap even at specified $\alpha$. The ambiguity is removed once
model parameters are fixed, as seen in Fig.\,3, but sensitivity to
$\alpha$ is well visible. The sensitivity indicates that, if we
are able to deduce values of $f$ from multi-color photometry, we
will have at hand a valuable new constraints on stellar
convection.

  \begin{figure}
  \centering
    \includegraphics[bb=63 281 400 733, width=88mm,clip]{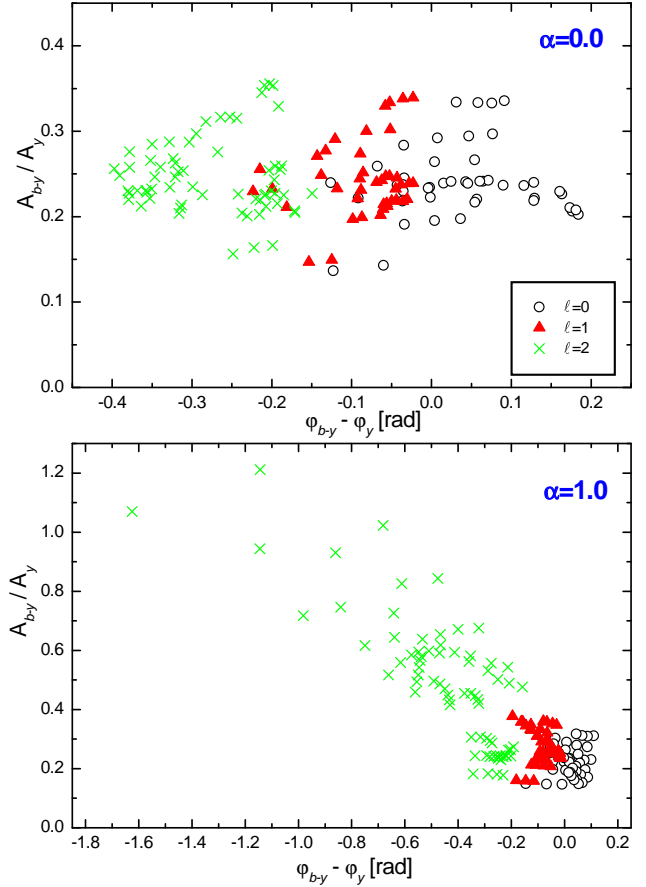}
             \caption{ The effect of mixing-length parameter on the
             locations of modes with different $\ell$ degree.
             Here we show the positions of $\ell=0,1,2$
             unstable modes in the diagnostic diagrams involving $b-y$
             and $y$ Str\"omgren filters for $\delta$ Scuti models
             of $1.9 M_\odot$. The frequency range covers $p_1$ to $p_4$
             radial modes. The upper and lower
             panels are for $\alpha=0$ and $\alpha=1.0$, respectively.}
        \label{aaaaaa}
         \end{figure}

 The test should be applied to more realistic modeling of convection
- pulsation interaction than used here. Our only aim here was to
show sensitivity of the nonadiabatic observables to convection and
we believe that our approximation is adequate for this aim. In section 4
we compare the calculated and measured $f$'s.

  \begin{figure}
  \centering
    \includegraphics[bb=63 281 400 733, width=88mm,clip]{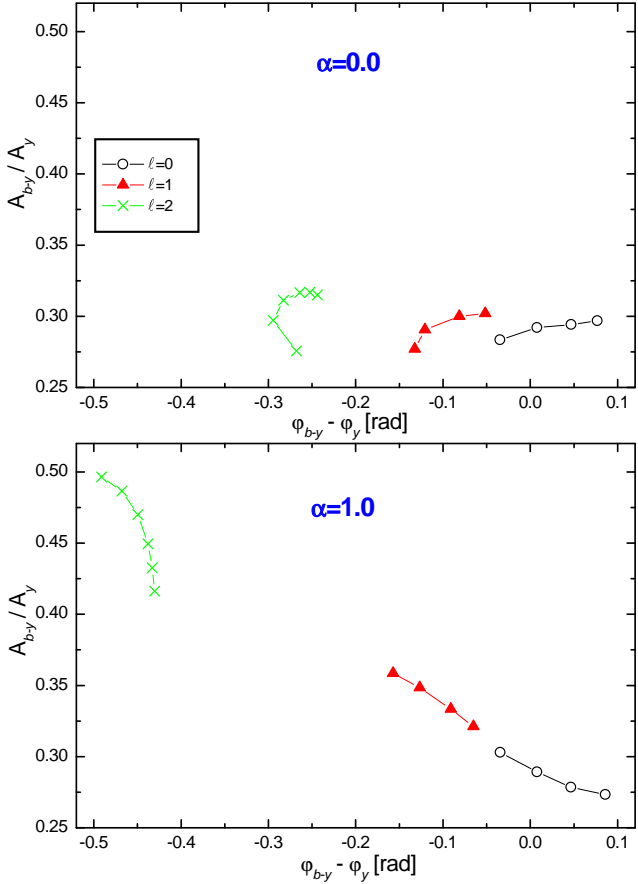}
             \caption{ The same as in Fig.\,2, but the one stellar model
              with $\log T_{\rm eff}=3.867$ is considered. The upper panel
              shows the photometric observables calculated at $\alpha=0.0$,
              and the lower one those at $\alpha=1.0$.}
        \label{aaaaaa}
         \end{figure}

\section{A method for inferring $f$ values from observations}

We begin with rewriting Eq.1 in the form of the following linear
equation.
$${\cal D}_{\ell}^{\lambda} ({\tilde\varepsilon} f) +{\cal
E}_{\ell}^{\lambda} {\tilde\varepsilon} =A^{\lambda}.\eqno(2)$$
where
$${\tilde\varepsilon}\equiv \varepsilon Y^m_{\ell}(i,0),$$
$${\cal D}_{\ell}^{\lambda}f\equiv b_{\ell}^{\lambda}
D_{1,\ell}^{\lambda},$$
$${\cal E}_{\ell}^{\lambda}\equiv
b_{\ell}^{\lambda}(D_{2,\ell}+D_{3,\ell}^{\lambda}).$$
Here the $\lambda$ superscript identifies the passband. Eqs.(2)
for a number of $\lambda$'s form a set of observational equations.
In the right-hand side we have measured amplitudes, $A^\lambda$,
expressed in the complex form. The quantities to be determined are
$({\tilde\varepsilon} f)$ and ${\tilde\varepsilon}$. Both must be
regarded complex. Of course we are primarily interested in the
value of $f$. However, inferred value of $\tilde\varepsilon$ may
also be useful as a constraint on mode identification if it is
found unacceptably large.
\begin{table*}
\begin{center}
\caption {Stellar parameters for three $\delta$ Sct stars:
$\beta$ Cas, 20 CVn and AB Cas.}
\begin{tabular}{|c|c|c|r|c|c|c|c|c|c|}
\hline
     Object  &   Sp  & Period [d] &  $\pi$ [mas]  & $\log T_{\rm eff}$ &  $\log L$  & $\log g*$ & $\log g**$ & [m/H] & $v_e\sin i$ \\
\hline
 $\beta$ Cas & F2III-IV & 0.1009 & $59.89\pm0.56$ & $3.856\pm0.01$ & $1.426\pm0.008$ & $3.60\pm0.01$ & 3.66 & 0.0 &   69   \\
\hline
     20 CVn  &   F3III  & 0.1217 & $11.39\pm0.69$ & $3.874\pm0.01$ & $1.881\pm0.053$ & $3.62\pm0.01$ & 3.49 & 0.5 &   15   \\
\hline
     AB Cas  &  A3V+KV  & 0.0583 &  $3.33\pm1.30$ & $3.908\pm0.01$ & $0.998\pm0.430$ & $4.55\pm0.01$ & 4.26 & 0.0 &   55   \\
\hline
\end{tabular}\\
\end{center}
{~~~~~~$*$  from photometry}
 {~~~~~~~$**$  from evolutionary tracks}
\end{table*}

Having data only for two passbands we can infer $f$ in a unique
way once we know $\ell$. However, we usually do not and therefore
we need at least three passband data. The procedure is to determine $f$
by means of $\chi^2$ minimization, assuming trial values of $\ell$.
We will regard the $\ell$ and the associated complex $f$ value as
the solution if it corresponds to $\chi^2$ minimum which is
significantly deeper than at other $\ell$'s. We will consider the
$\ell$ values up to six.

If we have data on spectral line variations, the set of equations
(2) may be supplemented with an expression relating
$\tilde\varepsilon$ to complex amplitudes of the first moments,
${\cal M}_1^{\lambda}$,
$$-{\rm i}\omega R \left( u_{\ell}^{\lambda} + \frac{GM
v_{\ell}^{\lambda} }{R^3\omega^2} \right) \tilde\varepsilon
={\cal M}_1^{\lambda}.\eqno(4)$$
where
$$u_{\ell}^{\lambda} = \int_0^1 h_\lambda(\mu) \mu^2
P_{\ell}(\mu) d\mu.$$
$$v_{\ell}^{\lambda} = \ell \int_0^1 h_\lambda(\mu) \mu
\left( P_{\ell-1}(\mu) - \mu P_{\ell}(\mu) \right)d\mu.$$
are coefficients representing a straightforward generalization
of the coefficients $u_{\ell}$ and $v_{\ell}$
introduced by Dziembowski (1977) for gray atmospheres.
We stress that the first moment is the only measure of the radial
velocity amplitude, which does not depend on the aspect and the
azimuthal number, $m$, just like the light amplitude.

There are uncertainties in model parameters, which enter
the expressions for $D_1^{\lambda}$ and $D_3^{\lambda}$. The
partial derivatives of ${\cal F}_\lambda |b_{\ell}^{\lambda}|$,
which appear there depend on $T_{\rm eff}$, $\log g$ and the
metallicity parameter $[m/H]$. We will not consider here the stars
with chemical peculiarities, thus we are adopting the solar
mixtures of heavy elements. We repeat our minimization for the
ranges of these three parameters consistent with observational
errors as well as with our evolutionary tracks.

If the best fit is for $\ell=0$, an additional constraint on models
follows from mode frequency, because the $\ell=0$ frequency
spectrum is sparse and the radial order of the mode may be easily identified.
In this case we can tune a star in the exact values
of the observed frequency. Stellar parameters have significant effect on
$\chi^2$. We will see that based on $\chi^2$ we can obtain more
stringent constrains on these parameters.

Generalization of the method to the case of modes coupled by
rotation is straightforward. Such modes are represented in terms
of a superposition of spherical harmonics with $\ell$'s differing
by 2 and the same $m$'s. The monochromatic amplitude of a coupled
mode is given by
$${\mathcal A}^{\lambda}(i)= \sum_k a_k A_{k}^{\lambda} (i)$$
where $a_k$ coefficients are solutions of the degenerate
perturbation theory for slowly rotating stars (see
Daszy\'nska-Daszkiewicz et al. 2002). The values of $a_k$
coefficients depend on the rotation rate and the frequency
distance between modes.

The counterpart of Eq.(2) is
$$\sum_k w_k^{\lambda} [ {\cal D}_{\ell_k}^{\lambda} (\varepsilon
f) + {\cal E}_{\ell_k}^{\lambda} \varepsilon]=A^\lambda,\eqno(5)$$
where

$$w_k^{\lambda}=Y^m_{\ell_k}(i,0) b_{\ell_k}^{\lambda} a_k.$$
The main difference relative to the single $\ell$ case is that now
the result depends on the aspect. Further, we expect a strong
dependence of calculated amplitude  on model parameters because
the values of $a_k$ are very sensitive to small frequency
distances between coupled modes.

\section{Applications}

We applied the method described above to data on three $\delta$
Sct variables: $\beta$ Cas, 20 CVn and AB Cas. In Table 1 we give
parameters for these stars. In this table we rely mostly on the
catalogue of Rodriguez et al. (2000).

The photometric data were dereddened according to Crawford (1979)
and Crawford \& Mandwewala (1976), and then the effective
temperatures, gravities and bolometric corrections were obtained
from Kurucz's (1998) tabular data. In Table 1 we give two values
of $\log g$, from Kurucz data and from evolutionary tracks. To
calculate the $D_1^{\lambda}$ and $D_3^{\lambda}$ coefficients we
took the second one, which we regard more reliable.
Errors in $\log T_{\rm eff}$ and $\log g$ correspond
to typical errors from the photometric calibration procedure.

For $\beta$ Cas and AB Cas we adopted solar metal abundance,
whereas for 20 CVn we used $[m/H]=0.5~(Z\approx0.06)$ due to Hauck
et al.(1985) and Rodriguez et al.(1998) and checked also the other
one, $[m/H]=0.3~(Z\approx0.04)$. The values of $\log L$ for $\beta$
Cas and 20 CVn are derived from the Hipparcos parallaxes. The
parallax for AB Cas is very inaccurate. The central value locates
this object close to the ZAMS. Though the star is a component of
an Algol-type binary system, it is reasonable to assume that it is
described by ordinary mass-conserving  models because it was
originally less massive and the episode of the rapid mass
accretion is most likely forgotten. This is what we assume in this
paper. Therefore, we adopted as the minimum luminosity the value
at ZAMS, and the maximum luminosity as the highest values allowed
by the Hipparcos parallax.
  \begin{figure*}
  \centering
    \includegraphics[bb=24 507 537 718, width=18cm,clip]{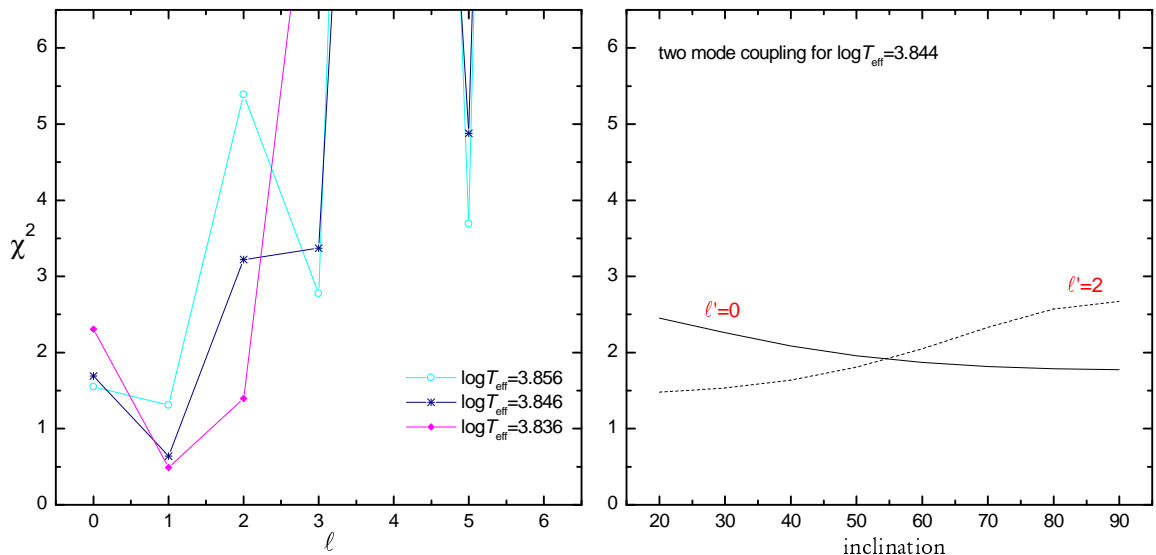}
             \caption{ In the left panel we show the $\chi^2$ as a function
         of $\ell$ obtained for three models of $\beta$ Cas star.
         In the right panel we plot the $\chi^2$ as a function of inclination
         considering the mode coupling between $\ell=0$ and $\ell=2$
         at $v_{\rm rot}=$70 km/s.}
        \label{aaaaaa}
         \end{figure*}

\subsection{$\beta$ Cas}

$\beta$ Cas is a $\delta$ Sct star with radial velocity variations
of 2 km/s (Mellor 1917). Mills (1966) classified it as $\delta$ Sct
variable on the basis of photometric observations.

$\beta$ Cas is one of a few $\delta$ Sct stars in which only one mode
has been detected so far. Rodriguez et al. (1992) identified this mode as
$p_2$ or $p_3$ of $\ell=1$, on the basis of photometric
diagrams in Str\"omgren filters. Balona \& Evers (1999), using
the same photometric data, did not get
an unambiguous identification.

We have rather precise parameters for this,
the brightest and the nearest $\delta$ Scuti variable.
Note, in particular, small luminosity errors in Tab.\,1.
The adopted metal abundance, $Z=0.02$, is based on the $[m/H]$ value obtained
from IUE spectra by Daszy\'nska \& Cugier (2002).
The photometric amplitudes and phases are from Rodriguez et al. (1992).

The frequency value combined with mean density implies that if the
mode were radial, it could be only $p_3$. We first assume that the
mode is adequately described in terms of single spherical harmonic
and consider $\ell$-values from 0 to 6. The uncertainties in the
values of $M$ and $R$ are inconsequential and we consider only
uncertainty in $T_{\rm eff}$. The estimated value of the mass is
$1.95 M_{\odot}$. In the left panel of Fig.\,4 we see that
the uncertainty on the effective temperature does not impair
identification of the $\ell$ value.
The $\chi^2$ minimum at $\ell=1$ is the deepest one, particularly
at the two lower effective temperatures.
  \begin{figure}
  \centering
    \includegraphics[bb=72 420 450 733, width=88mm,clip]{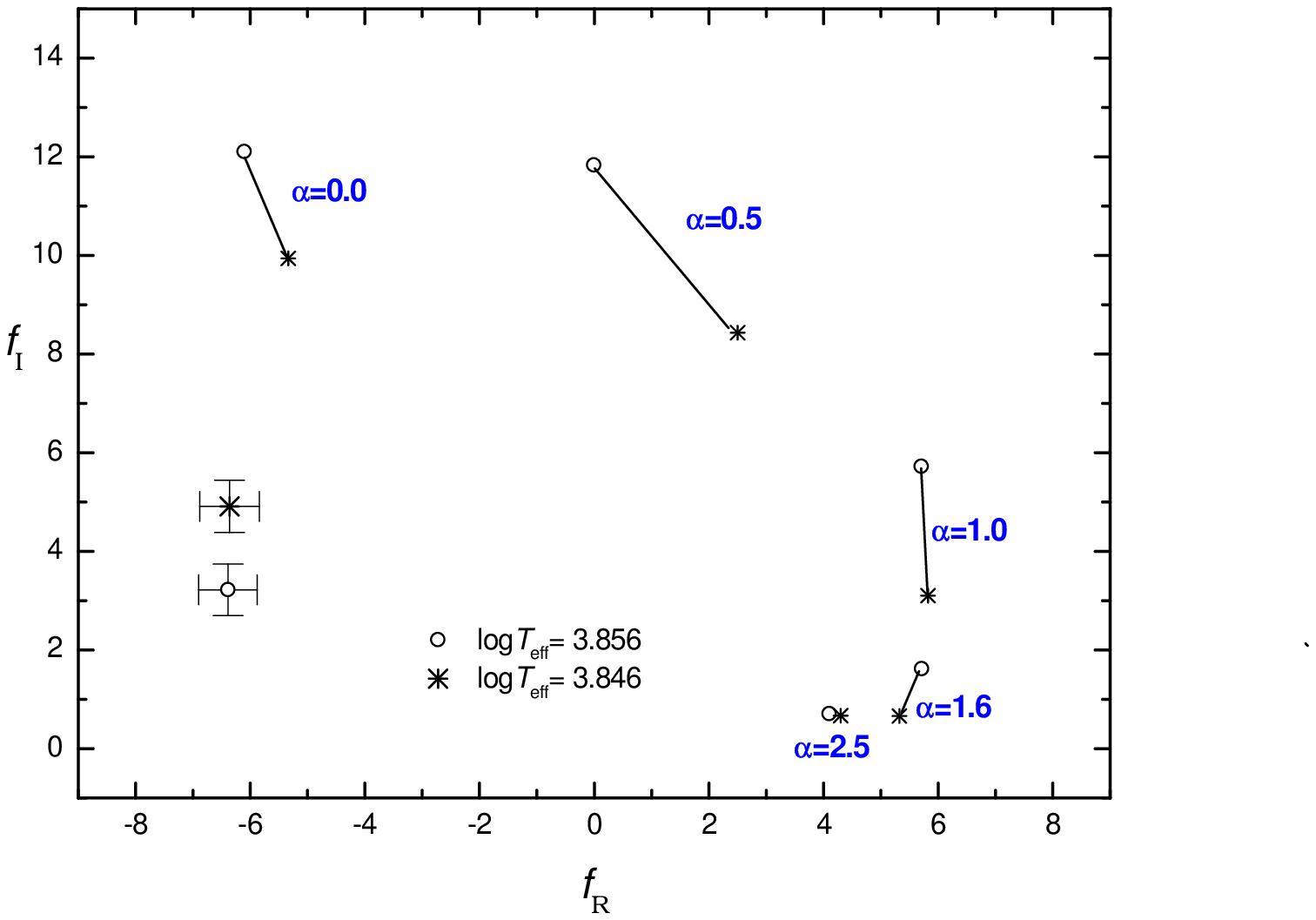}
        \caption{
        Comparison of the $f$ values inferred from Str\"omgren photometry
        for $\beta$ Cas with the values from the models calculated with various
        values of the MLT parameter $\alpha$ and the mass of $1.95 M_{\odot}$.
        Points with the error bars represent inferred values assuming
        two indicated effective temperatures.
       }
        \label{aaaaaa}
         \end{figure}

The star is a relatively rapid rotator, $v_{\rm rot}$ is at least 70 km/s,
therefore we have to consider the possibility that the mode is a
coupled one, most likely an $\ell=0$ and $\ell=2$ superposition
(see Daszy\'nska-Daszkiewicz at al. 2002).
Relying on the formalism described at the end of Section 3,
we evaluated the best $f$-values and assiociated $\chi^2$
as a function of the aspect.
In the right panel we show $\chi^2$ variation with the
inclination. We can see that at no value of the inclination
the $\chi^2$ is nearly as low as at $\ell=1$. Thus we conclude that
the mode excited in $\beta$ Cas is most likely a single $\ell=1$ mode.

Values of $f$ corresponding to this identification are shown in
Fig.\,5 together with the error bars,
which are the errors from the least-square method.
The uncertainty in
temperature is more significant than the errors of $f$
determination. The theoretical $f$ values were calculated for
$M=1.95 M_{\odot}$ assuming five values of the mixing-length
parameter $\alpha$: 0.0, 0.5, 1.0, 1.6 and 2.5.  Still the constraint
on convection is interesting because the range of the acceptable
values of $f$ is narrower than the range of the calculated values
with different $\alpha$'s.

The observed values of $f_R$ are closer to those calculated
with $\alpha=0$, which may be taken as an evidence that convection
in the H ionization zone is relatively inefficient.
However, values of $f_I$ require rather higher $\alpha$'s (about 1)
instead. In view of our crude treatment of convection--pulsation
interaction we have to take these indications with a great caution.

  \begin{figure}
  \centering
    \includegraphics[bb=62 442 450 733, width=88mm,clip]{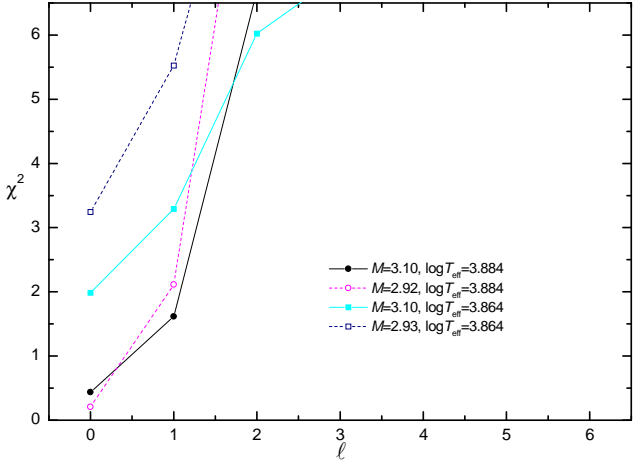}
         \caption{ Dependence of $\chi^2$ on $\ell$ for four models of 20 CVn
            on the edges of the error box obtained for $Z=0.06$.}
        \label{aaaaaa}
         \end{figure}
  \begin{figure}
  \centering
    \includegraphics[bb=52 398 480 733, width=88mm,clip]{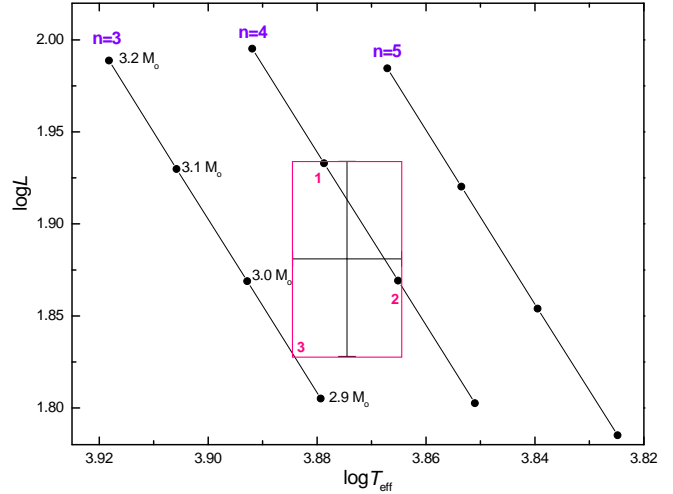}
         \caption{ The observational error box for 20 CVn on HR diagram.
           The lines of constant radial order, $n=3,4,5$, are also drawn.}
        \label{aaaaaa}
         \end{figure}
  \begin{figure}
  \centering
    \includegraphics[bb=62 458 412 730, width=88mm,clip]{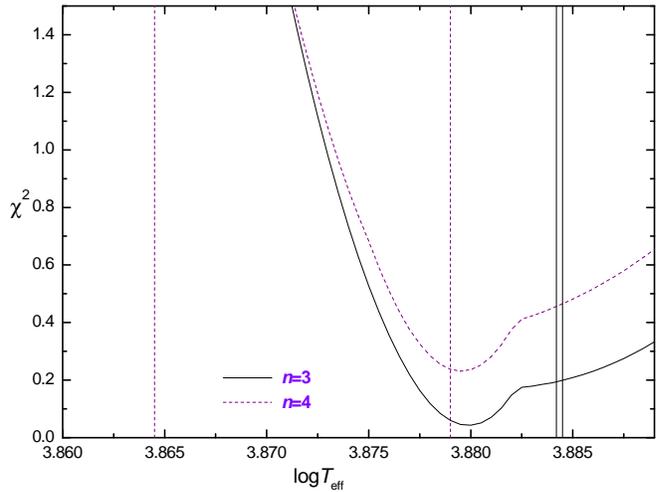}
         \caption{ The variation of $\chi^2$ with effective temperature
           for the models along the lines $n=3$ and $n=4$.}
        \label{aaaaaa}
         \end{figure}

  \begin{figure}
  \centering
    \includegraphics[bb=72 420 450 733, width=88mm,clip]{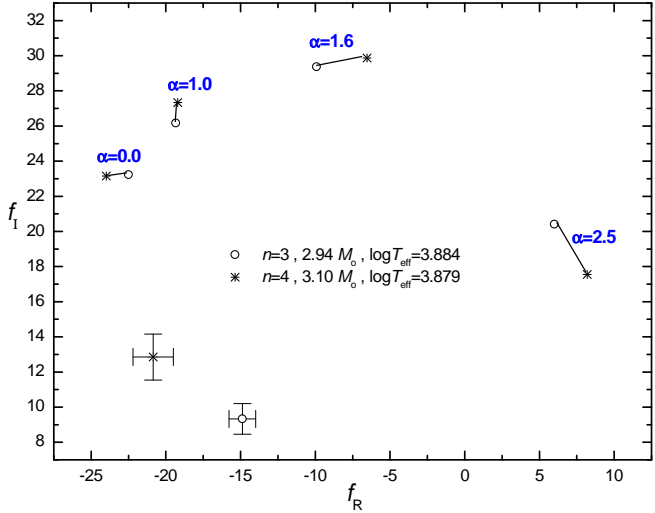}
         \caption{ The same as in Fig.\,5, but with empirical values of $f$
           for the models with min$\chi^2$ for $n=3$ and $n=4$ of 20 CVn.}
        \label{aaaaaa}
         \end{figure}

\subsection{20 CVn}

This $\delta$ Scuti variable is also regarded to be monoperiodic
(e.g. Shaw 1976, Pe{\~n}a \& Gonzalez-Bedolla 1981). The mode was
identifited as $\ell=0$ by means of photometry (Rodriguez et al.
1998) as well as spectroscopy (Chadid et al. 2001). Our result
shown in Fig.\,6 for $Z=0.06$ clearly confirms the previous
identification. For models on the edges of the error box the
minimum of $\chi^2$ is at $\ell=0$. The same is true for $Z=0.04$,
but the values of the $\chi^2$ are higher.

As we mentioned in Sect.\,2, having such identification of
pulsating mode we can refine stellar parameters by fitting the
observed period. Still, we have to consider various radial orders,
$n$. In Fig.\,7 we show the HR diagram with the error box
representing uncertainty of stellar parameters and the lines of
the constant period ($P$ = 0.1217 d) for $n=3,4,5$, obtained with
evolutionary models calculated for $Z=0.06$ and indicated masses.
Only models along these lines are allowed. We can see that as for
the radial order we have only two possibilities, $n=3$ or $n=4$.
In the case of $Z=0.04$ only $n=4$ is allowed but the $\chi^2$ is
significantly larger. We thus see that our method allows to
constrain the $Z$ value.

In Fig.8 we show the variations of the $\chi^2$ with $\log T_{\rm
eff}$ for the tuned models with $n=3$ and $n=4$. The vertical
lines correspond to the intersections of the constant period lines
with the error box, so that only the values of $\log T_{\rm eff}$
between these lines for a given $n$ are allowed.
The models yielding the lowest $\chi^2$ are just models 1 and 3 in Fig.7
for $n=4$ and $n=3$, respectively.
The values of $\chi^2$ are 0.19 for $n=3$ and 0.24 for $n=4$, what
indicates that both are possible as identifications of the radial
order in this star.

Fig.\,9 shows the empirical values of the nonadiabatic $f$-parameter
for models corresponding to the deepest $\chi^2$ minima for four
values of $\alpha$: 0.0, 1.0, 1.6 and 2.5.
Relative positions of empirical and theoretical $f$ values in 20 CVn
are qualitatively similar to those in $\beta$ Cas (cf. Fig.\,5).
However, these values, both calculated and inferred,
are considerably higher than in the case of $\beta$ Cas,
which is a consequence of higher radial order.  The $\ell=1$ mode
identified in $\beta$ Cas has frequency between the $n=2$ and
$n=3$ radial modes.

\subsection{AB Cas}

The next example is the primary component of an Algol-type system.
As we have already pointed out, luminosity of this star is very
uncertain due to the large error in the parallax.
We will see that with our method we can
significantly improve the accuracy of the stellar parameters as
provided that the identified mode is radial.
  \begin{figure}
  \centering
    \includegraphics[bb=62 435 450 733, width=88mm,clip]{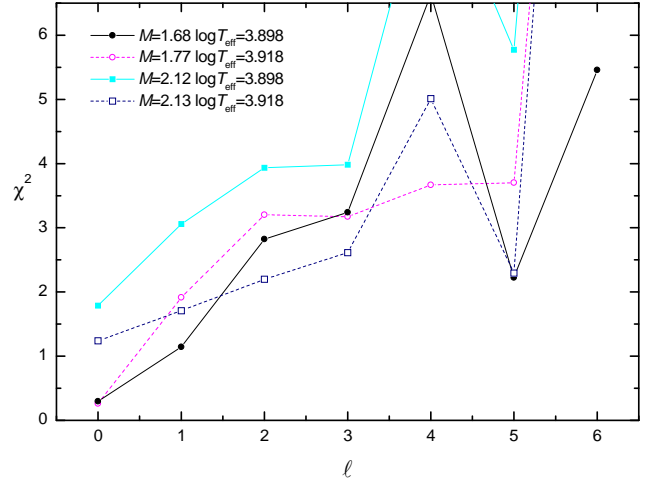}
         \caption{ Dependence of $\chi^2$ on $\ell$ for four models
         on the edges of the error box of AB Cas.}
\label{aaaaaa}
\end{figure}
\begin{figure}
\centering
\includegraphics[bb=62 396 480 733, width=88mm,clip]{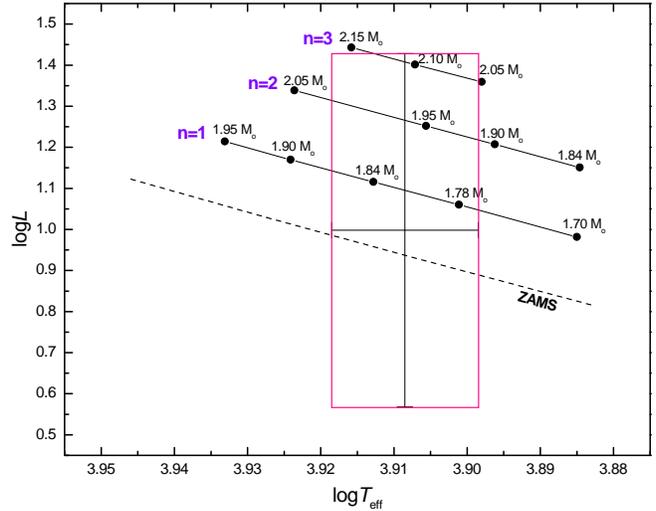}
\caption{ The observational error box for AB Cas on
HR diagram.  The lines of constant radial order with $n=1,2,3$ are also shown.}
\label{aaaaaa}
\end{figure} %

  \begin{figure}
  \centering
    \includegraphics[bb=72 420 450 733, width=88mm,clip]{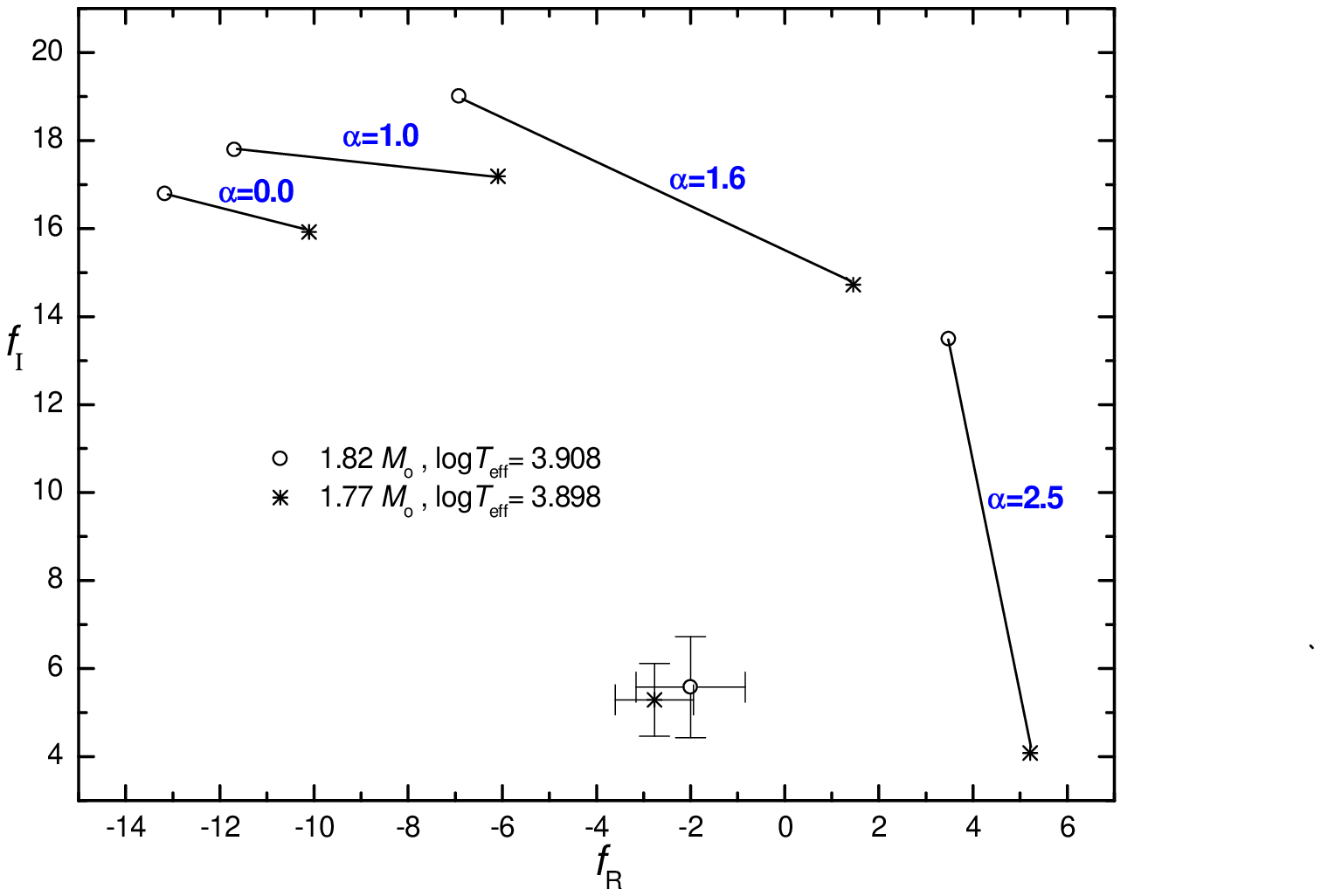}
         \caption{ The same as in Fig.\,5, but with empirical values of $f$ for
    two tuned models of AB Cas with  $n=1$ and $\log T_{\rm eff}=3.908$ and 3.898.}
        \label{aaaaaa}
         \end{figure}

In the whole range of allowed parameters $\chi^2$ has by far the
deepest minimum at $\ell=0$, as we can see in Fig.\,10. This
identification is in agreement with that of Rodriguez et al.
(1998). In Fig.\,11 we show the HR diagram with the error box
representing uncertainty of stellar parameters and the lines of
the constant period ($P$ = 0.0583 d) for $n=1,2,3$, obtained with
evolutionary models calculated for $Z=0.02$ and indicated masses.
The deepest minimum of $\chi^2$ (0.25) along the $n=1$ line occurs
at $M=1.77, \log T_{\rm eff}=3.8985, \log L=1.047$. The lowest
value of $\chi^2$ for $n=2$ (0.62) is at $M=1.91, \log T_{\rm
eff}=3.8985, \log L=1.217$, and for $n=3$ we got $\chi^2=1.02$,
$T_{\rm eff}=3.8985, \log L=1.361$. We can see that $n=1$
identification is strongly favored. For such mode identification
the value of $\log L$ is between 1.047 and 1.144 leading to
$\chi^2=0.25$ and 0.56, respectively.

In Fig.\,12 we compare empirical values of $f$ with ones
calculated for various mixing-length parameter, $\alpha$. The star
is the hottest of the three objects. Still, calculated values are
strongly affected by the choice of $\alpha$.  Just like the
previous, regardless the value of $\alpha$, the empirical values
of the real and imaginary parts of $f$, could not be
simultaneously reproduced with our model calculations.

\section{Conclusions and discussion}

We believe that the three examples of $\delta$ Scuti stars
considered in the previous section, clearly show that there is a
wider application of multicolor data on excited mode than explored
so far. In addition to the spherical harmonic degree, $\ell$,
we were able with our $\chi^2$ minimization method to infer
the value of the complex parameter $f$ which describes the
bolometric flux perturbation and yields a strong constraint on
models of stellar convection. In the two cases when determined
value of $\ell$ was zero, we were able to determine the radial
order of the mode and significantly limit the uncertainty of the
stellar parameters.

The values of $f$ we found could not be reproduced with model
calculations made with the convective-flux-freezing approximation
for any value of the MLT parameter $\alpha$.
We should not be surprised. After all, the approximation
adopted is grossly inadequate. Though there are certain
common features in the relative positions of the deduced
and calculated $f$'s in the three cases considered, it is
premature to draw any conclusions about properties
of stellar convection from this fact.

Inadequacy in our treatment of convection is not the only possible
cause of the large discrepancy between the calculated and empirical
$f$'s. Our use of the Eddington approximation in
calculation of the nonadiabatic oscillations may be an
oversimplification. We believe it is of secondary importance as
the value of $f$ is determined at the optical depth $\tau\gg1$.
Our use of static atmospheric models with the depth-independent $g$
seems also a good approximation. Still, it should be kept that these
two approximation should be at some point verified. The accuracy of
the atmospheric models is of a greater concern, which was discussed
recently by Heiter et al.(2002), where sensitivity of the atmospheric
structure and observable quantities to the convection treatment
was demonstrated (see also Smalley \& Kupka 1997). Whatever is
the cause of discrepancy, the goal in model calculations should
be to achieve consistent $f$'s.

Determination of $\ell$ is an independent goal. It is important
that it could be done without an {\it a priori} knowledge of $f$.
This is not a new finding. This has been done earlier for both
$\delta$ Scuti and $\beta$ Cephei variables. In particular,
in the latter case, when $f$ may be approximately treated as real, the three
color data often allow for an unambiguous $\ell$ determination
(e.g. Heynderickx et al. 1994, Balona\& Evers 1999 ). For
$\delta$ Scuti stars the situation is more complicated. Although, we
believe, our method represents an improvement, it still
does not always work . We may believe in the inferred values of
both $f$ and $\ell$ only if the minima of $\chi^2$ are strongly
$\ell$ dependent. This was the true in all the three cases
considered in the previous section. It is not always so. For the
$\delta$ Scuti star 1 Mon, for instance, we found $\chi^2$ quite
flat between $\ell=0$ and 2. For this star we also made an
unsuccessful attempt to use the data of the radial velocity as
provided by Balona et al.(2001), but even including this quantity
did not change the result. This failure should not discourage
future efforts to combine spectroscopic data.

The main message of our work is that applications of multi-color photometry
data of $\delta$ Scuti stars go beyond identification of the
$\ell$ values of the excited  mode. We showed that such data allow to refine
parameters of the stars and, what we regard most important, yield
strong constraints on models of stellar convection.

\acknowledgements{ The work was supported by KBN grant No. 5 P03D
012 20. }

\end{document}